\documentclass[preprint,12pt]{elsarticle}

\usepackage{amssymb}
\usepackage{amsmath}
\usepackage{multirow} 

\usepackage{lineno}

\journal{Nuclear Physics B}

\begin{document}

\begin{frontmatter}



\title{\boldmath The branching fraction measurements of $J/\psi$ decay into $\rho \eta$ and $\phi \eta$ final states}


\author[a]{V.\,V.\,Anashin}
\author[a]{O.\,V.\,Anchugov}
\author[a]{K.\,V.\,Astrelina}
\author[a,b]{V.\,M.\,Aulchenko}
\author[a]{V.\,V.\,Balakin}
\author[a,b]{E.\,M.\,Baldin}
\author[a,c,d]{G.\,N.\,Baranov}
\author[a]{A.\,K.\,Barladyan}
\author[a,d]{A.\,Yu.\,Barnyakov}
\author[a]{M.\,Yu.\,Barnyakov}
\author[a]{I.\,Yu.\,Basok}
\author[a,b]{E.\,A.\,Bekhtenev}
\author[a,b]{O.\,V.\,Belikov}
\author[a]{D.\,E.\,Berkaev}
\author[a,b]{A.\,E.\,Blinov}
\author[a,b,d]{V.\,E.\,Blinov}
\author[a,b]{A.\,V.\,Bobrov}
\author[a,b]{V.\,S.\,Bobrovnikov}
\author[a,b,c]{A.\,V.\,Bogomyagkov}
\author[a]{D.\,Yu.\,Bolkhovityanov}
\author[a,b]{A.\,E.\,Bondar}
\author[a,d]{V.\,M.\,Borin}
\author[a,b]{A.\,R.\,Buzykaev}
\author[a,c]{V.\,L.\,Dorokhov}
\author[a,c]{D.\,V.\,Dorokhova}
\author[a,b]{F.\,A.\,Emanov}
\author[a,d]{D.\,N.\,Grigoriev}
\author[a,b]{V.\,V.\,Kaminskiy}
\author[a,c]{S.\,E.\,Karnaev}
\author[a]{G.\,V.\,Karpov}
\author[a,c,d]{K.\,Yu.\,Karukina}
\author[a,d]{P.\,V.\,Kasyanenko}
\author[a]{A.\,A.\,Katcin}
\author[a,b]{T.\,A.\,Kharlamova}
\author[a]{V.\,A.\,Kiselev}
\author[a]{A.\,A.\,Kovalenko}
\author[a,b]{E.\,V.\,Kozyrev}
\author[a,c]{A.\,A.\,Krasnov}
\author[a,b]{E.\,A.\,Kravchenko}
\author[a,b]{V.\,N.\,Kudryavtsev}
\author[a,b]{V.\,F.\,Kulikov}
\author[a,d]{G.\,Ya.\,Kurkin}
\author[a]{I.\,A.\,Kuyanov}
\author[a,b]{D.\,A.\,Kyshtymov\corref{cor1}}
\ead{D.A.Kyshtymov@inp.nsk.su}  
\author[a]{N.\,N.\,Lebedev}
\author[a,c,d]{E.\,B.\,Levichev}
\author[a]{P.\,V.\,Logachev}
\author[a,b]{D.\,A.\,Maksimov}
\author[a]{T.\,V.\,Maltsev} 
\author[a]{V.\,M.\,Malyshev}
\author[a,c]{R.\,Z.\,Mamutov}
\author[a,b]{A.\,L.\,Maslennikov}
\author[a,b]{O.\,I.\,Meshkov}
\author[a,b]{I.\,I.\,Morozov}
\author[a,c]{I.\,A.\,Morozov} 
\author[a]{A.\,A.\,Murasev}
\author[a]{S.\,A.\,Nikitin}
\author[a,b]{I.\,B.\,Nikolaev}
\author[a,c]{I.\,N.\,Okunev}
\author[a]{S.\,B.\,Oreshkin}
\author[a,b]{A.\,A.\,Osipov}
\author[a,b]{I.\,V.\,Ovtin}
\author[a]{A.\,V.\,Pavlenko}
\author[a,b]{S.\,V.\,Peleganchuk}
\author[a]{K.\,G.\,Petrukhin}
\author[a,c]{P.\,A.\,Piminov}
\author[a,d]{S.\,G.\,Pivovarov}
\author[a]{A.\,V.\,Polyanskiy}
\author[a,b]{V.\,G.\,Prisekin}
\author[a,b]{O.\,L.\,Rezanova}
\author[a,b]{A.\,A.\,Ruban}
\author[a]{\framebox[1.06\width][s]{G.\,A.\,Savinov}}
\author[a,b,d]{D.\,V.\,Senkov}
\author[a,b]{A.\,G.\,Shamov} 
\author[a,b]{L.\,I.\,Shekhtman}
\author[a]{D.\,A.\,Shvedov}
\author[a,b]{B.\,A.\,Shwartz}
\author[a]{E.\,A.\,Simonov}
\author[a]{S.\,V.\,Sinyatkin}
\author[a,c]{M.\,A.\,Skamarokha}
\author[a]{A.\,N.\,Skrinsky}
\author[a,b]{A.\,V.\,Sokolov}
\author[a,b]{E.\,V.\,Starostina}
\author[a,b,d]{D.\,P.\,Sukhanov}
\author[a,b]{A.\,M.\,Sukharev}
\author[a,b]{A.\,A.\,Talyshev}
\author[a,b]{V.\,A.\,Tayursky}
\author[a,b]{V.\,I.\,Telnov}
\author[a,b]{Yu.\,A.\,Tikhonov}
\author[a,b]{K.\,Yu.\,Todyshev}
\author[a]{Yu.\,V.\,Usov}
\author[a]{A.\,I.\,Vorobiov}
\author[a,b]{V.\,N.\,Zhilich}
\author[a,c]{A.\,A.\,Zhukov}
\author[a,b]{V.\,V.\,Zhulanov}
\author[a,c]{A.\,N.\,Zhuravlev}
\author[a,c]{D.\,A.\,Zubkov}


\affiliation[a]{organization={Budker Institute of Nuclear Physics},
            addressline={akademika Lavrentieva prospect 11}, 
            city={Novosibirsk},
            postcode={630090}, 
            country={Russia}}
\affiliation[b]{organization={Novosibirsk State University},
            addressline={Pirogova street 1}, 
            city={Novosibirsk},
            postcode={630090}, 
            country={Russia}}
\affiliation[c]{organization={Synchrotron Radiation Facility - Siberian Circular Photon Source "SKIF"},
            addressline={Morskoy prospect 2}, 
            city={Novosibirsk},
            postcode={630090}, 
            country={Russia}}

\affiliation[d]{organization={Novosibirsk State Technical University},
            addressline={Karl Marx prospect 20}, 
            city={Novosibirsk},
            postcode={630090}, 
            country={Russia}}

\cortext[cor1]{Corresponding author at: Budker Institute of Nuclear Physics, Novosibirsk, 630090, Russia.}

\begin{abstract}
We present measurements of the branching fractions for the $J/\psi$ meson decays into the $\rho\eta$ and $\phi\eta$ final states, based on data collected with the KEDR detector at the VEPP-4M collider. The data set consisted of 4.93 million $J/\psi$ events. The resulting branching fractions are:
\begin{itemize}
    \item $\mathcal{B}(J/\psi \to \rho\eta) = (2.04 \pm 0.58 \pm 0.39)\times 10^{-4}$,
    \item $\mathcal{B}(J/\psi \to \phi\eta) = (7.82 \pm 1.17 \pm 0.58) \times 10^{-4}$,
\end{itemize}
where the first uncertainty is statistical and the second one is systematic. 

In the study of the $\rho\eta$ decay, the dynamics of $J/\psi\to\pi^+\pi^-\eta$ is analyzed. The hints from contributions of $\rho(1450)\eta$ and $a^{\pm}_2\pi^{\mp}$ intermediate states are observed. Additional measured branching fractions in the model that includes $\rho(1450)\eta$ and $a^{\pm}_2\pi^{\mp}$ are:
\begin{itemize}
    \item $\mathcal{B}(J/\psi \to \pi^+\pi^-\eta) = (4.73 \pm 0.49 \pm 1.17)\times10^{-4}$,
    \item  $\mathcal{B}(J/\psi \to (a_2^+\pi^- + a_2^- \pi^+)) = (1.05\pm 0.37 \pm 0.34)\times 10^{-3}$,
    \item  $\mathcal{B}(J/\psi \to \rho(1450)\eta \to \pi^+\pi^-\eta) < 1.51 \times 10^{-4}$, at a confidence level of 90\%.
\end{itemize}
All results are consistent with previous studies.
\end{abstract}




\end{frontmatter}




\section{Introduction}
\label{sec:intro}


The $J/\psi$ meson, discovered in 1974, remains one of the most important tools for studying quantum chromodynamics (QCD) in the charm quark sector. As the lightest vector charmonium state, its decays provide unique insights into both perturbative and non-perturbative aspects of QCD. Despite nearly five decades of study, many aspects of $J/\psi$ decays, particularly those involving light hadrons in the final state, are not yet fully understood.


The decays of $J/\psi$ into vector-pseudoscalar ($V+P$) final states, such as $\rho\eta$ and $\phi\eta$, are particularly interesting as a decay class. They proceed through strong (three-gluon) and electromagnetic (single-photon) mechanisms, providing a testing ground for understanding their relative contributions.



The $\rho\eta$ channel was first studied in the late 1980s by MARK-III \cite{MARK-III} and DM2 \cite{DM2}. This channel is the dominant process in $J/\psi \to \pi^+ \pi^- \eta$, accounting for approximately half of all events. This decay is especially complex due to several factors: significant $\rho$-$\omega$ interference effects (despite the small branching fraction $\omega\to\pi^+\pi^-$) possible contributions from higher mass resonances like $\rho(1450)$ and additional processes such as $a_2^{\pm}\pi^{\mp}$ decays. While BaBar \cite{BaBar_piPpiMeta} and BES-III \cite{Bes-III} have studied $J/\psi \to \pi^+ \pi^- \eta$ decay, the complete dynamical description, including contributions from $\rho(1450)\eta$ (analogous to the $\pi^+\pi^-\pi^0$ case \cite{BaBar_3pi}) and $a^{\pm}_2\pi^{\mp}$, is still lacking.

In contrast, the $\phi\eta$ channel offers a cleaner experimental signature due to the narrow width of the $\phi$ meson and absence of significant interference effects. This decay has been studied by multiple experiments including MARK-III \cite{MARK-III}, DM2 \cite{DM2}, BaBar \cite{BaBar_phi}, Belle \cite{Belle} and BES-II \cite{Bes-II_phi}, with generally consistent results but varying precision.

Currently, the KEDR detector \cite{KEDR} of the VEPP-4M collider \cite{VEPP4M} accumulated a sample of 4.93 million $J/\psi$ decays. This dataset enables a new study of both the $\rho \eta$ and $\phi\eta$ decay channels with a more detailed dynamic analysis of the final states.

\section{Theoretical framework}

\subsection{$\pi^+\pi^-\eta$ final state}
For the $\pi^+\pi^-\eta$ channel, we have to consider:
\begin{itemize}
\item $\rho-\omega$ interference, despite the small branching fraction $\omega\to\pi^+\pi^-$. The contributions from $\omega\eta$ and $\rho\eta$ are comparable due to isospin suppression in $J/\psi\to\rho\eta$,
\item Additional contributions from intermediate states:
\begin{itemize}
\item $\rho(1450)\eta$,
\item $a^{\pm}_2\pi^{\mp}$.
\end{itemize}
\end{itemize}

The differential cross section for vector-pseudoscalar ($V+P$) decays can be expressed as:

\begin{multline}
\label{eq:VP_cross_section}
\left(\frac{d\sigma}{d\Gamma}\right)_{V+P} \propto \left| \sum_{k=\rho, \omega, \rho(1450)} a_k e^{i\phi_k} \right|^2 = \sum_{k=\rho, \omega, \rho(1450)} |a_k|^2 \\
+ \sum_{k,j=\rho, \omega, \rho(1450), k \neq j} a_k a_j^* e^{i(\phi_k-\phi_j)},
\end{multline}
where the resonance amplitude $a_{\rho}$, written for $\rho$ meson, takes the form:
\begin{multline}
\label{eq:rho_amplitude}
a_{\rho} = |\mathbf{P_{\pi^+}} \times \mathbf{P_{\pi^-}}| \sin(\theta_n) \frac{m^2_\rho}{q^2 - m^2_\rho + iq\Gamma_\rho(q^2)} \\
\times \sqrt{\mathcal{B}{(J/\psi\to\rho\eta)}\mathcal{B}{(\rho\to\pi\pi)}},
\end{multline}
with the energy-dependent width:

\begin{equation}
\label{eq:rho_width}
\Gamma_\rho(q^2) = \Gamma_\rho\left(\frac{p_\pi(q^2)}{p_\pi(m_\rho^2)}\right)^3 \left(\frac{m_\rho^2}{q^2}\right).
\end{equation}
Here:
\begin{itemize}
\item $|\mathbf{P_{\pi^\pm}}|$ are charged pion momenta,
\item $\theta_n$ is the angle between the normal to the reaction plane and the beam axis,
\item $m_\rho$ and $\Gamma_\rho$ are the $\rho$ meson mass and width,
\item $q$ is the pion pair invariant mass,
\item $p_{\pi}$ is the pion momentum in the $\rho$ rest frame.
\end{itemize}

The same matrix element has been used in the \cite{SND_matrix_element}.

The $\rho-\omega$ interference terms in Eq.~\ref{eq:VP_cross_section} can be expressed as:

\begin{equation}
\label{eq:function_interf}
\begin{gathered}
a_{\rho} a_{\omega}^* e^{-i\phi} + a_{\rho}^* a_{\omega} e^{i\phi} =  \\
\frac{\sqrt{\mathcal{B}(J/\psi \to \rho \eta) \mathcal{B}(\rho \to \pi^+ \pi^-)}\sqrt{\mathcal{B}(J/\psi \to \omega \eta) \mathcal{B}(\omega \to \pi^+ \pi^-)}}{((q^2-m^2_{\rho})^2+q^2\Gamma^2_{\rho}(q^2))((q^2-m^2_{\omega})^2+q^2\Gamma^2_{\omega}(q^2))} \\
\times 2|\mathbf{P_{\pi^+}} \times \mathbf{P_{\pi^-}}|^2 sin^2(\theta_n)  m^2_{\rho} m^2_{\omega}\\
\times\Bigg[ (q^4+m^2_{\rho}m^2_{\omega}+q^2\Gamma_{\rho}(q^2)\Gamma_{\omega}(q^2))cos(\phi) \\
- q^2(m^2_{\rho}+m^2_{\omega})cos(\phi) \\
+ (q^3\Gamma_{\omega}(q^2)+q\Gamma_{\rho}(q^2)m^2_{\omega})sin(\phi) \\
- (q^3\Gamma_{\rho}(q^2)+q\Gamma_{\omega}(q^2)m^2_{\rho})sin(\phi) \Bigg],
\end{gathered}
\end{equation}
where $\phi$ is the interference phase between resonances. All other interference terms follow similar expressions, with phases defined relative to the $\rho$ meson.

The total cross section includes the $a^{\pm}_2\pi^{\mp}$, which are tensor-pseudoscalar ($T+P$) decays and can't be described by Eq.~\ref{eq:VP_cross_section}. Therefore, the total cross section for both vector-pseudoscalars and tensor-pseudoscalars without interference effects between them is: 


\begin{equation}
\label{eq:total_cross_section}
\left(\frac{d\sigma}{d\Gamma}\right)_{\pi^+\pi^-\eta} = \left(\frac{d\sigma}{d\Gamma}\right)_{V+P} + \left(\frac{d\sigma}{d\Gamma}\right)_{T+P}.
\end{equation}

\subsection{$K^+K^-\eta$ final state}
For the $\phi\eta$ channel studied through $K^+K^-\eta$, we consider:
\begin{equation}
\label{eq:phi_cross_section}
\left(\frac{d\sigma}{d\Gamma}\right)_{\phi\eta} \propto |a_\phi|^2,
\end{equation}
where $a_{\phi}$ follows Eq.~\ref{eq:rho_amplitude} with $\rho\to\phi$ and $\pi\to K$ replacement. 

The total cross section includes contributions from other intermediate states with $M_{K^+K^-}$ much bigger than the $\phi$ meson mass. Therefore, all other contributions, except $\phi$ resonance, will be called non resonant contributions:

\begin{equation}
\label{eq:KKeta_total}
\left(\frac{d\sigma}{d\Gamma}\right)_{K^+K^-\eta} = \left(\frac{d\sigma}{d\Gamma}\right)_{\phi\eta} + \left(\frac{d\sigma}{d\Gamma}\right)_{\text{non res}} + \left(\frac{d\sigma}{d\Gamma}\right)_{interference~term}.
\end{equation}

\section{Simulation of $J/\psi$ decays}
\label{sec:simulation}

The Monte Carlo simulation incorporates several components:

\begin{itemize}
\item The vector-pseudoscalar decays (Eq.~\ref{eq:VP_cross_section}--\ref{eq:rho_width}) were simulated using the generator from~\cite{rhopi},
\item The tensor-pseudoscalar channel $a_2^{\pm}\pi^{\mp}$ (without interference effects) and the most dominant background process $\pi^+\pi^-\pi^0\gamma\gamma$ for $\pi^+\pi^-\eta$ final state were simulated using a  generator developed for the~\cite{malyshev} work, which is based on helicity formalism  \cite{helicity_1,helicity_2},
\item non resonant $J/\psi\to K^+K^-\eta$ decays and continuum events from $e^+e^-\to\pi^+\pi^-\eta$ were simulated using generator from \cite{rhopi} and formula from Eq.~\ref{eq:rho_amplitude} with infinite decay width of vector particle. 
\end{itemize}

The simulation of $J/\psi \to\pi^+ \pi^- \pi^0 \gamma\gamma$ has been done considering that it proceeded through $\omega\eta$, $\phi\eta$, $a_2\rho$ intermediate states, as well as through non resonant states. The weights between these contributions were tuned to agree with the $\pi^+\pi^-\pi^0$ invariant mass distribution in the $\pi^+ \pi^- \pi^0 \gamma\gamma$ decay, when the invariant mass of photons is near the $\eta$ meson ($450$ MeV $< M_{\gamma\gamma} < 650$ MeV).

Each process was generated using particle properties from PDG~\cite{PDG}. To simulate other background events such as $\rho \pi$, $K \pi K_s$, $K \pi K$, $K^+K^-\pi^0\pi^0$, $f_0(500)\gamma$, $\omega\pi^0$ the BES generator \cite{BES_generator} and the generators from \cite{rhopi, malyshev} are used. Parameters of JETSET \cite{jetset} used in the BES generator were tuned in the work~\cite{KEDR_experiment}.

\section{Event selection}
\label{sec:events_sel}


The analysis is performed using the KEDR detector at the VEPP-4M collider. A detailed description of the detector can be found in \cite{KEDR}. The data used include 1.4 pb$^{-1}$ at the $J/\psi$ peak and 0.278 pb$^{-1}$ outside the peak. In total, approximately 4.93 million $J/\psi$ decays are analyzed. The data outside the $J/\psi$ peak are used to estimate the continuum background events.

The signature used for the analysis is two central tracks and two photons originating from the $\eta$ meson. These photons are clusters in the calorimeter with an energy above 50 MeV that are not associated with charged tracks. The central tracks have to cross a cylinder with a radius of 3 cm and a length of 17 cm around the interaction point. Any number of non-central tracks is allowed. Events with more than two photons are also used for systematic uncertainty estimation.
 
Kinematic fit is performed assuming the hypotheses of $\pi^+\pi^-\eta$ and $K^+K^-\eta$, which is five-constraint kinematic fit. Since this work does not use an identification system, a $\chi^2$ value of less than 70 is used as a criterion to separate events involving pions and kaons. 

To select events containing the $\eta$ meson and suppress background, each event is reprocessed in the hypothesis of free photons. An invariant mass of these photons ($M_{\gamma\gamma}$) has to be between 500 MeV and 600 MeV.

The $\pi^+\pi^-\eta$ analysis employs additional selection criteria:
\begin{itemize}
\item $\chi^2_{\pi^+\pi^-\eta} < \chi^2_{K^+K^-\eta}$ (to suppress $K^+K^-\eta$ background),
\item $cos(\theta_{\gamma\gamma})>0.4$ (to suppress $\rho\pi$ background),
\end{itemize}
where $\theta_{\gamma\gamma}$ is the angle between photons. The total selection efficiency for $\rho\eta$ process is 17\%.

For the $K^+K^-\eta$ process, an additional selection criteria beyond the baseline requirements:
\begin{itemize}
\item $\chi^2_{K^+K^-\eta} < \chi^2_{\pi^+\pi^-\eta}$ (to suppress $\pi^+\pi^-\eta$ background),
\item $M_{K^+ K^-} < 1090$ MeV (to analyze only $\phi$ resonance).
\end{itemize}

Unlike the $\pi^+\pi^-\eta$ case, no angular selection is required. The total selection efficiency for $\phi\eta$ process is 9\%.

The most dominant background events for $\pi^+\pi^-\eta$ decay come from $\pi^+\pi^-\pi^0\gamma\gamma$ process despite the strict criterion on the number of photons. Also, the continuum events from $e^+e^-\to\pi^+\pi^-\eta$ contribute to the final state. Outside the $J/\psi$ peak, there are two background events registered, which agree with the BaBar cross-section continuum measurements \cite{BaBar_piPpiMeta}. Using the BaBar cross-section result and integrated luminosity in the $J/\psi$ peak, the continuum contribution has been fixed in the current work.

The $K^+K^-\eta$ system is cleaner; there are no events outside the $J/\psi$ meson.

\section{Analysis procedure}
\label{sec:analysis}

\subsection{$\pi^+\pi^-\eta$ final state}

The $\pi^+\pi^-\eta$ final state is analyzed using a Dalitz plot distribution of invariant masses between $\pi^+\eta$ and $\pi^{-}\eta$. 

Following event generation (Section~\ref{sec:simulation}), we construct invariant mass histograms $H_i[j, l]$ for selected events (Section~\ref{sec:events_sel}) for each $i$-th cross section component, where indexes $j$ and $l$ correspond to the bins in the $M_{\pi^+\eta}$ and $M_{\pi^{-}\eta}$ projections of the Dalitz plot distribution. The event counts in these histograms are proportional to the simulated integral of luminosity. For interference terms between decays $k$ and $m$ (Eq.~\ref{eq:function_interf}), we define:

\begin{itemize}
\item $H_{k,m}^{c\pm}$: Cosine interference terms,
\item $H_{k,m}^{s\pm}$: Sine interference terms.
\end{itemize}

The expected event distribution is parameterized as:

\begin{equation}
\label{eq:function_n_exp}
\begin{gathered}
N_{exp}[j, l] = \Bigg(p_{1}H_{\rho\eta}[j, l] + p_{2}H_{\omega\eta}[j,l] +  p_{3}H_{\rho(1450)\eta}[j,l] \\
+\sqrt{p_1p_2}\Bigg[(H^{c+}_{\rho\eta,\omega\eta}[j,l]-H^{c-}_{\rho\eta,\omega\eta}[j,l])cos(\phi_{\rho,\omega}) \\
+ (H^{s+}_{\rho\eta,\omega\eta}[j,l]-H^{s-}_{\rho\eta,\omega\eta}[j,l])sin(\phi_{\rho,\omega})\Bigg] \\
+\sqrt{p_1p_3}\Bigg[(H^{c+}_{\rho\eta,\rho(1450)\eta}[j,l]-H^{c-}_{\rho\eta,\rho(1450)\eta}[j,l])cos(\phi_{\rho,\rho(1450)}) \\
+ (H^{s+}_{\rho\eta,\rho(1450)\eta}[j,l]-H^{s-}_{\rho\eta,\rho(1450)\eta}[j,l])sin(\phi_{\rho,\rho(1450)})\Bigg] \\
+\sqrt{p_2p_3}\Bigg[(H^{c+}_{\rho(1450)\eta, \omega\eta}[j,l]-H^{c-}_{\rho(1450)\eta, \omega\eta}[j,l])cos(\phi_{\rho,\omega} - \phi_{\rho,\rho(1450)}) \\
+(H^{s+}_{\rho(1450)\eta, \omega\eta}[j, l]-H^{s-}_{\rho(1450)\eta, \omega\eta}[j,l])sin(\phi_{\rho,\omega}- \phi_{\rho,\rho(1450)})\Bigg] \\
+ p_{4}H_{a^{\pm}_2\pi^{\mp}}[j,l] + p_{5}H_{\pi^+\pi^-\pi^0\gamma\gamma}[j,l] \\ + H_{cont. + interf.}[j,l] \Bigg),        
\end{gathered}
\end{equation}
where $p_i$ and $\phi_k$ are resonance strengths and relative phases, $H_i[j,l]$ is the invariant mass histogram for the selected events in the $[j, l]$ bin. The $p_i$ are proportional to the branching fractions. Here, a $J/\psi\to\pi^+\pi^-\pi^0\gamma\gamma$ is considered as it is the dominant background for this process. The continuum events including the interference terms are taken into account in the $H_{cont. + interf.}[j,l]$ histogram with free phase relative to $\rho\eta$ process. Interference effects are considered only with $V+P$ decays.

Parameters $p_i$ and $\phi_k$ are determined by maximizing the likelihood function \cite{likelihood}:
\begin{equation}
\label{eq:likelihood}
L_{\pi^+\pi^-\eta} = -2\sum_{j,l}\left(N_{exp}[j,l] - N_{obs}[j,l] + N_{obs}[j,l]\ln\frac{N_{obs}[j,l]}{N_{exp}[j,l]}\right),
\end{equation}
summed over all histogram bins, where $N_{obs}[j,l]$ are experimentally observed events in the $[j,l]$ bin.


Using $\mathcal{B}(J/\psi\to\omega\eta)$ from BES-II measurement~\cite{Bes-II_omega} as a constraint, we fit data using Eq.~\ref{eq:likelihood} and calculate:

\begin{equation}
\begin{split}
&\mathcal{B}(J/\psi\to\rho\eta)=\frac{p_1I_{\rho\eta}}{N_{J/\psi}\mathcal{B}(\eta\to\gamma\gamma)\epsilon_{\rho\eta}}, \\
&\mathcal{B}(J/\psi\to\rho(1450)\eta\to\pi^+\pi^-\eta)=\frac{p_3I_{\rho(1450)\eta}}{N_{J/\psi}\mathcal{B}(\eta\to\gamma\gamma)\epsilon_{\rho(1450)\eta}}, \\
&\mathcal{B}(J/\psi\to a^{\pm}_2\pi^{\mp})=  \frac{p_4I_{a^{\pm}_2\pi^{\mp}}}{N_{J/\psi}\mathcal{B}(\eta\to\gamma\gamma)\mathcal{B}(a^{\pm}_2\to\pi^{\pm}\eta)\epsilon_{a^{\pm}_2\pi^{\mp}}},\\
&\mathcal{B}(J/\psi\to \pi^+\pi^-\eta)=\frac{N_{\pi^+\pi^-\eta}}{N_{J/\psi}\mathcal{B}(\eta\to\gamma\gamma)\epsilon_{\pi^+\pi^-\eta}},\\    
\end{split}
\label{eq:br_formula}
\end{equation}
where $N_{J/\psi}$ --- amount of the $J/\psi$ mesons, $\mathcal{B}(\eta\to\gamma\gamma)$ and $\mathcal{B}(a^{\pm}_2\to\pi^{\pm}\eta)$ --- branching fractions of $\eta$ decay into $\gamma \gamma$ and $a^{\pm}_2$ decay into $\pi^{\pm}\eta$, $I_i$ is a number of events in the $H_i$ histogram, $\epsilon_i$ --- selection efficiency for $i$-th process (see Eq. \ref{eq:efficiency_1} - \ref{eq:efficiency_2}).

\begin{equation}
\label{eq:efficiency_1}
\epsilon_i = n_i/n_{total},
\end{equation}
where $n_i$ and $n_{total}$ are the numbers of selected and total events in the simulation for $i$-th process. For $\mathcal{B}(J/\psi\to \pi^+\pi^-\eta)$ case, the average efficiency over all contributions in the decay is taken:
\begin{equation}
\label{eq:efficiency_2}
\epsilon_{\pi^+\pi^-\eta} = \frac{\sum_i N_i\times\epsilon_i}{\sum_i N_i},
\end{equation}
the summation is carried out over all channels in the $\pi^+\pi^-\eta$ decay including interference terms (see Eq. \ref{eq:function_n_exp}), $N_i$ is a number of events of each term determined after the fit procedure.

Number of the $\pi^+\pi^-\eta$ events is:
\begin{equation}
N_{\pi^+\pi^-\eta} = (N_{obs}-N_{\pi^+\pi^-\pi^0\gamma\gamma} - N_{cont}),
\end{equation}
where $N_{\pi^+\pi^-\pi^0\gamma\gamma}=p_4I_{\pi^+\pi^-\pi^0\gamma\gamma}$ is a number of $\pi^+\pi^-\pi^0\gamma\gamma$ events and $N_{cont}$ --- continuum events including interferences.

\subsection{$K^+K^-\eta$ final state}
For $J/\psi\to\phi\eta\to K^+K^-\eta$, we use a simplified one dimensional model:

\begin{equation}
\begin{gathered}
N_{exp}[j] = p_6H_{\phi\eta}[j] + p_7H_{non~res}[j] \\
+\sqrt{p_6p_7}\Bigg[(H^{c+}_{\phi\eta,non~res}[j]-H^{c-}_{\phi\eta,non~res}[j])cos(\phi_{\phi\eta,non~res}) \\
+ (H^{s+}_{\phi\eta,non~res}[j]-H^{s-}_{\phi\eta,non~res}[j])sin(\phi_{\phi\eta,non~res})\Bigg],
\end{gathered}
\end{equation}
where $\phi_{\phi\eta,non~res}$ is a free phase. The branching fraction is:
\begin{equation}
\mathcal{B}(J/\psi\to\phi\eta) = \frac{p_6I_{\phi\eta}}{N_{J/\psi}\mathcal{B}(\eta\to\gamma\gamma)\mathcal{B}(\phi\to K^+K^-)\epsilon_{\phi\eta}}.
\end{equation}

\section{Experimental results}

The analysis uses the following branching fractions:
\begin{itemize}
\item $\mathcal{B}(\rho\to\pi^+\pi^-)=100\%$,
\item $\mathcal{B}(a_2^{\pm}\to\pi^{\pm}\eta)=(14.5\pm1.2)\%$,
\item $\mathcal{B}(\eta\to\gamma\gamma)=(39.36\pm0.18)\%$,
\item $\mathcal{B}(\phi\to K^+K^-)=(49.1\pm0.5)\%$.
\end{itemize}

All values are taken from PDG~\cite{PDG}.

\subsection{Results for $\rho\eta$ channel}
\label{sec:rhoeta_results}

\begin{figure*}[hbtp]
\parbox{0.5\textwidth}{\centering \includegraphics [width=0.5\textwidth]{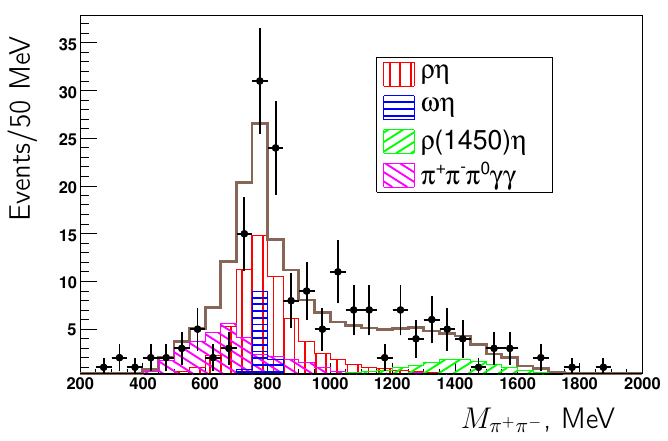}
}
  \hfill
  \parbox{0.5\textwidth}{\centering \includegraphics [width=0.5\textwidth]{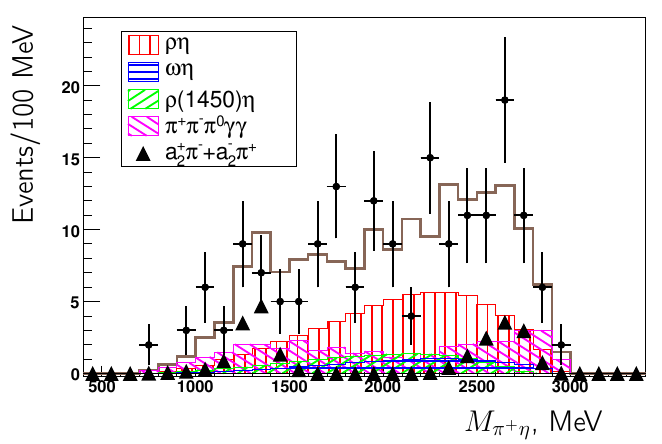}
}
 \\
  \parbox{0.5\textwidth}{\centering (a)}
  \hfill
  \parbox{0.5\textwidth}{\centering (b)}
   \\
  \parbox{\textwidth}{
\caption{
The (a) $M_{\pi^+\pi^-}$ and (b) $M_{\pi^{+}\eta}$ distributions. The Kolmogorov probability of agreement for $M_{\pi^+\pi^-}$ is 0.43 and for $M_{\pi^{+}\eta}$ is $\sim$1. Here, the black dots are observed events in the experiment. The result distribution is shown as a brown line, that includes all interference effects between $\rho\eta$, $\omega\eta$ and $\rho(1450)\eta$ processes. Others individual contributions without interferences are: red histogram is the $\rho\eta$ process, blue --- $\omega\eta$, green --- $\rho(1450)\eta$, pink --- $\pi^+\pi^-\pi^0\gamma\gamma$, black triangles --- $a^{\pm}_2\pi^{\mp}$. In the $M_{\pi^+\pi^-}$ picture the contribution from $a^{\pm}_2\pi^{\mp}$ is not shown due to limited space in the figure.}
\label{fig:mpippimk}}
 \end{figure*}

Figure~\ref{fig:mpippimk} presents the (a) $M_{\pi^+\pi^-}$ and (b) $M_{\pi^{+}\eta}$ distributions after the Dalitz plot fit, $L_{\pi^+\pi^-\eta}/n.d.o.f. = 207/227$ --- is a ratio of the likelihood function and the number of degrees of freedom.

The figures illustrate the potential contributions from the decays of $\rho (1450)\eta$ and $a^{\pm}_2\pi^{\mp}$ to the $\pi^ + \pi ^ - \eta$ state. The fit procedure without these intermediate states gives worse likelihood function values. The interference phases of $\rho-\omega$, $\rho-\rho(1450)$ and $\rho$ - continuum  with only statistical uncertainty are equal to $(88.4 \pm 2.5)$, $(148.6\pm 44.3)$ and $(47.0 \pm 161.7)$ degrees, respectively.


From the fit procedure the total 134 events from $\pi^+\pi^-\eta$ decay have been measured in which there are 67 events --- $\rho\eta$, 12 --- $\omega\eta$ (fixed), 16 --- $\rho(1450)\eta$, 23 --- $a_2^{\pm}\pi^{\mp}$. The continuum events are equal to 4; other events come from interferences. The obtained branching fractions according to our model are:

\begin{align*}
\mathcal{B}(J/\psi\to\rho\eta)=(1.99\pm0.58)\times 10^{-4}, \\
\mathcal{B}(J/\psi\to \pi^+\pi^-\eta)=(4.63\pm0.49)\times 10^{-4},\\
\mathcal{B}(J/\psi\to a^{\pm}_2\pi^{\mp})=(1.03\pm0.37)\times 10^{-3},\\
\mathcal{B}(J/\psi\to\rho(1450)\eta\to\pi^+\pi^-\eta)=(6.23\pm4.70)\times 10^{-5},
\end{align*}
where only statistical uncertainties are presented.

The measured branching fraction for $a^{\pm}_2\pi^{\mp}$ is lower than a theoretical phenomenological estimation ($3.8\times 10^{-3}$) \cite{a2pi} and is consistent with an upper limit from experimental measurement ($4.3\times 10^{-3}$) \cite{a2pi_exp}.

\subsection{Systematic uncertainty analysis for $\rho\eta$ decay}
\label{sec:systematics}

The systematic uncertainty from event selection is evaluated by varying the criteria as shown in Table~\ref{Tab:var_rho}.

	\begin{table*}[hbtp]
 \centering
    \begin{tabular}{l c c c}
 	\hline
    \multirow{2}{*}{Criterion} &\multirow{2}{*}{Selection}&\multirow{2}{*}{Range variation}& Variation of \\ 
    & & & $\mathcal{B}(J/\psi\to\rho\eta)$ in \% \\\hline
    $\chi^2_{\pi^+\pi^-\eta}$ &  $\chi^2_{\pi^+\pi^-\eta} < 70$ &  $\chi^2_{\pi^+\pi^-\eta} < 90$ & $4.6$  \\  
    $\chi^2_{K^+K^-\eta}$ & $\chi^2_{\pi^+\pi^-\eta} < \chi^2_{K^+K^-\eta}$ & $\chi^2_{\pi^+\pi^-\eta} < 0.8\chi^2_{K^+K^-\eta}$  &  negligible  \\ 	 
    $cos(\theta_{\gamma \gamma})$  & $0.4 < cos(\theta_{\gamma \gamma})$ & $0.5 < cos(\theta_{\gamma \gamma})$ & $1.6$ \\  
    $M_{\gamma \gamma}$, MeV & $500 < M_{\gamma \gamma} < 600$ & $470 < M_{\gamma \gamma} < 630$ & $8.5$ \\  \hline
     \multicolumn{3}{c}{Sum in quadrature} & $9.8$ \\ \hline

    \end{tabular}
    \caption{\label{Tab:var_rho} The selection criteria variations for $\rho \eta$ process. In the fourth column the $J/\psi\to\rho\eta$ branching fraction variation from the main result is shown.}
\end{table*}

\begin{figure*}[hbtp]
\parbox{0.5\textwidth}{\centering \includegraphics [width=0.5\textwidth]{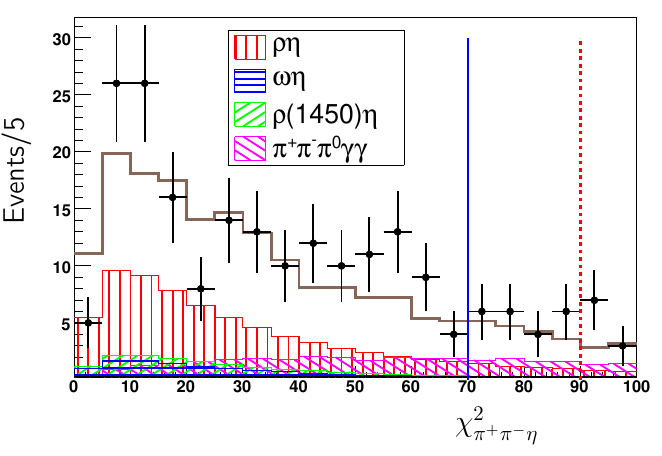}
}
  \hfill
  \parbox{0.5\textwidth}{\centering \includegraphics [width=0.5\textwidth]{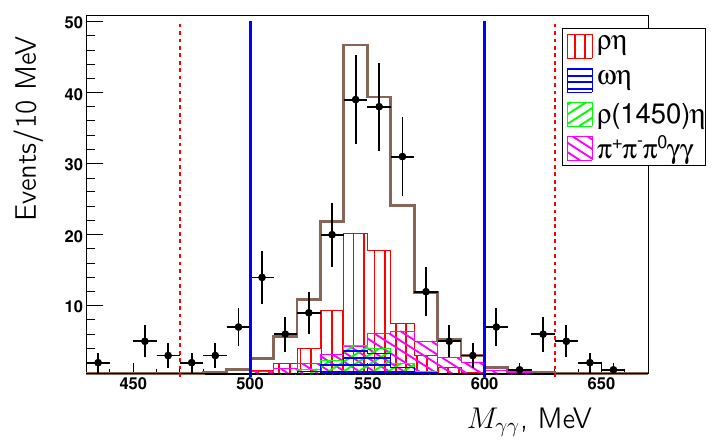}
}
 \\ 
 \vspace{\baselineskip}
 \\
  \parbox{0.5\textwidth}{\centering (a)}
  \hfill
  \parbox{0.5\textwidth}{\centering (b)}
   \\
  \parbox{\textwidth}{\caption{\label{fig:cuts} The variations of selection criteria for (a) $\chi^2_{\pi^+\pi^-\eta}$ and (b) $M_{\gamma\gamma}$. The applied selection criterion is shown as the solid blue line, its variation as the red dotted line. The Kolmogorov probability of agreement for these distributions for selected events are (a) 0.74 and (b) 0.50. The black dots are observed events in the experiment. The result distribution is shown as a brown line, red histogram is the $\rho \eta$ process, blue --- $\omega \eta$, green --- $\rho(1450) \eta$, pink --- $\pi^+ \pi^- \pi^0  \gamma\gamma$.}}
 \end{figure*}

Figure~\ref{fig:cuts} demonstrates the variation of:
\begin{itemize}
\item[(a)] $\chi^2_{\pi^+\pi^-\eta}$ criterion, where it can be seen that background events from $\pi^+\pi^-\pi^0\gamma\gamma$ become larger than $\pi^+\pi^-\eta$ at $\chi^2_{\pi^+\pi^-\eta}>70$,
\item[(b)] $M_{\gamma\gamma}$ distribution with wider mass window,
\end{itemize}
Here, the solid blue line corresponds to the applied selection criterion, and the red dotted line --- its variation.

In the experiment, two clusters of energy greater than 50 MeV were used in the calorimeter to identify the $\eta$ meson. To estimate the systematic errors associated with determining the number of photons, the photon number was increased to three, and the pair with the lowest $\chi^2$ value of the kinematic fit was selected. The energy of the additional photon was required to be less than 100 MeV. The resulting uncertainty is around 10.5\%. 

The influences of the track inefficiency reconstruction and the photon loss for the process $\pi^+\pi^-\pi^0$ were studied in \cite{rhopi}. A correction for the branching fraction of $2.2\%$ was derived with uncertainties due to track inefficiency reconstruction of $\pm 0.5\%$ and photon loss of $\pm 0.2\%$. The same correction is applied to $\pi^+\pi^-\eta$ data; therefore, all branching fractions have to be increased by 2.2\%. The momentum and angular resolution uncertainties have been estimated by two methods. The first one is scaling a drift chamber spatial resolution, and the second one is scaling systematic uncertainties of the tracking system calibration parameters for the simulated events. The difference of 3\% between results has been taken as the systematic uncertainty. Therefore, the total detector related uncertainty is 3.1\% (see~Table \ref{Tab:rec_ef}). The contribution of the nuclear interaction has been estimated as 1.4\% by comparison of the FLUKA \cite{fluka} and GHEISHA \cite{gheisha} packages in the simulation. The number of $J/\psi$ mesons is known with an accuracy of 1.1\% \cite{rhopi}. 

 \begin{table}[hbtp]
 \centering
     \begin{tabular}{l c}
 	\hline
        Source & Uncertainty, \% \\ \hline
        Track inefficiency reconstruction & 0.5 \\   
        Photon loss & 0.2 \\   
        Tracking $p/\theta$ resolution & 3.0\\ 	 \hline
        Sum in quadrature & 3.1 \\  \hline
    \end{tabular}
    \caption{\label{Tab:rec_ef} Detector related systematic uncertainty.}
\end{table}

The unaccounted interference effects between tensor mesons and between $V+P$ and $T+P$ processes contribute to the measurement by 10.7\%. This was estimated by comparing the results of fitting models that included interferences under zero degrees between $a_2^{\pm}\pi^{\mp}$ and with $\rho\eta$, as it is the dominant signal, and excluded interferences. The shift in the results of branching fraction measurements is the estimation of systematic uncertainty due to the interference effects.

An uncertainty from the fixation of $\rho$ meson parameters such as mass and width during the simulation was evaluated as 0.8\%. To estimate an uncertainty contribution from the constraint at the $\mathcal{B}(J/\psi \to \omega\eta)$, the $\mathcal{B}(J/\psi \to \omega\eta)$ has been varied up and down by the value of its uncertainty in the \cite{Bes-II_omega}. A larger $\rho\eta$ branching fraction shift from these variations is taken as an estimation of uncertainty. The same has been done with $\mathcal{B}(\omega\to \pi^+\pi^-)$ constraint. The total constraint uncertainty from the $\omega$ meson is 2.6\%. The statistical contribution from the simulated events, associated with the $\epsilon_{\rho\eta}$, is 0.3\%. 

The non resonant $e^+e^-\to\pi^+\pi^-\eta$ contribution uncertainty is found to be 0.9\%. It was determined by increasing its amount of events at the range of measured cross section uncertainty in the~\cite{BaBar_piPpiMeta} and comparing the results. 

To estimate the uncertainty associated with the simulation of the $J/\psi \to \pi^+ \pi^- \pi^0\gamma\gamma$, each weight in the simulation has been varied according to the scale of  branching fractions relative uncertainty \cite{PDG}. The shift of the branching fraction measurements due to the use of the varied weights of $J/\psi \to\pi^+ \pi^- \pi^0 \gamma\gamma$ process is taken as uncertainty estimation and corresponds to 2.7\%.

In addition to $\pi^+ \pi^- \pi^0 \gamma\gamma$, the uncertainty from other physical background events such as $\phi \eta$, $\rho \pi$, $K \pi K_s$, $K \pi K$, $K^+K^-\pi^0\pi^0$, $f_0(500)\gamma$ has been estimated. Their contributions to the final result have been taken into account in the fit procedure using known branching fractions from \cite{PDG} and selection efficiencies. A systematic uncertainty contribution from the physical background is 4.7\%. 

In the Table \ref{Tab:total_usr_rho_eta} all systematic uncertainties are presented.

 \begin{table}[hbtp]
 \centering
     \begin{tabular}{l c}
 	\hline
      Source & Uncertainty, \% \\ \hline
      Detector related uncertainty & 3.1 \\  
      Nuclear interaction & 1.4 \\ 
      Number of $J/\psi$ & 1.1\\ 	
      $m_{\rho}$ and $\Gamma_{\rho}$ simulation & 0.8 \\ 
      $\mathcal{B}(J/\psi \to \omega\eta)\times \mathcal{B}(\omega \to \pi^+\pi^-)$ & 2.6 \\   
      $\mathcal{B}(\eta\to\gamma\gamma)$  & 0.5 \\ 
      Selection efficiency ($\epsilon_{\rho\eta}$) & 0.3 \\ 
      Non res. $e^+e^-\to\pi^+\pi^-\eta$  & 0.9 \\ 
      $J/\psi \to \pi^+ \pi^- \pi^0 \gamma\gamma$ simulation & 2.7 \\   
      Physical background  & 4.7 \\  
      Interference effects of $a_2^{\pm}\pi^{\mp}$ & 10.7\\ 	
      Fixation of the photon number & 10.5\\ 	
      Selection criteria & 9.8 \\   \hline
      Total systematic uncertainty & 19.3 \\  \hline
    \end{tabular}
    \caption{\label{Tab:total_usr_rho_eta} The overall systematic uncertainty in the $\rho \eta$ process.}
\end{table}

\subsection{Results for $\phi\eta$ channel}
\label{subsec:phi_eta}

\begin{figure*}[hbtp]
\parbox{1.\textwidth}{\centering \includegraphics [width=0.9\textwidth]{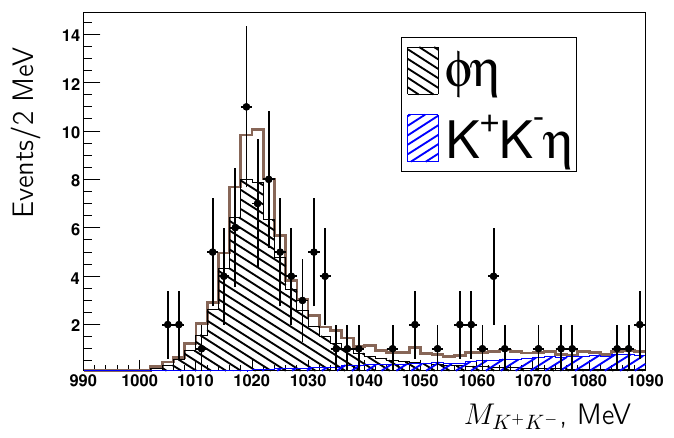}
}
  \parbox{\textwidth}{\caption{\label{fig:mkpkmk} The data fit result for the $M_{K^+K^-}$ invariant mass obtained by maximizing the likelihood function $L_{K^+K^-\eta}$. The Kolmogorov probability of agreement for this distribution is $\sim$1. The observed events from the experiment are shown as black dots in the figure. The resulting distribution is presented as a brown line. The black histogram represents the contribution from the $\phi \eta$ channel, while the blue one --- the non resonant $J/\psi \rightarrow K^+K^-\eta$ decay.}}
 \end{figure*}

Figure~\ref{fig:mkpkmk} shows the $K^+K^-$ invariant mass distribution, fitted using the $\phi\eta$ resonant component and non resonant $K^+K^-\eta$ contribution. The ratio of the likelihood function over the number of degrees of freedom is 46/50.

From the fit procedure there are 64 $\phi\eta$ events have been measured. The interference phase between $\phi$ meson and non resonant contribution is equal to $(-6.0 \pm 9.1)$ degrees, where uncertainty is statistical. The measured branching fraction is:
\begin{eqnarray}
\mathcal{B}(J/\psi\to\phi\eta) = (7.65\pm1.17)\times10^{-4},
\end{eqnarray}
where only statistical uncertainty is presented.

\subsection{Systematic uncertainty analysis for $\phi\eta$ decay}

\begin{figure*}[hbtp]
\parbox{0.5\textwidth}{\centering \includegraphics [width=0.5\textwidth]{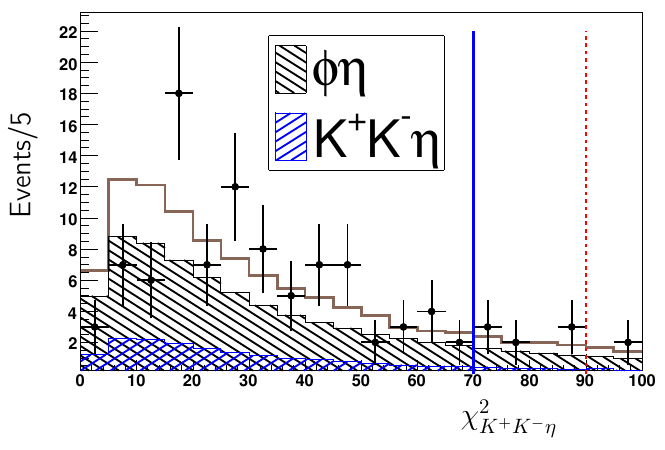}
}
  \hfill
  \parbox{0.5\textwidth}{\centering \includegraphics [width=0.5\textwidth]{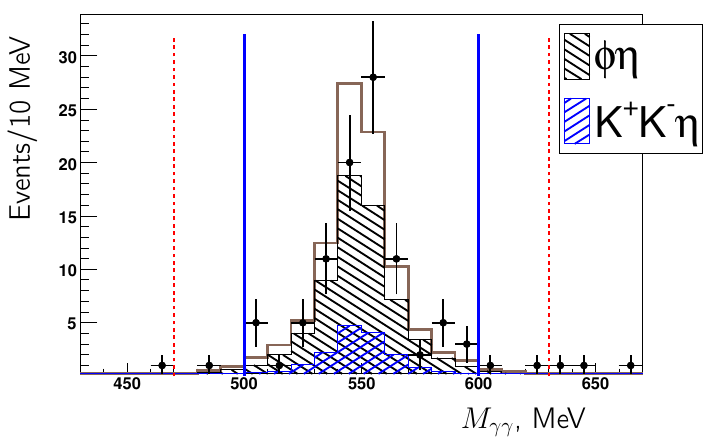}
}
 \\ 
 \vspace{\baselineskip}
 \\
  \parbox{0.5\textwidth}{\centering (a)}
  \hfill
  \parbox{0.5\textwidth}{\centering (b)}
   \\
  \parbox{\textwidth}{\caption{\label{fig:cuts_2} The variations of selection criteria for (a) $\chi^2_{K^+K^-\eta}$ and (b) $M_{\gamma\gamma}$. The solid blue line represents the used selection criteria and red dotted line --- variations. The Kolmogorov probability of agreement for these distributions for selected events are (a) 0.21 and (b) 0.87. The black dots --- observed events, brown line --- resulting distribution, black histogram --- $\phi \eta$ process, blue --- non resonant $K^+K^-\eta$.}}
 \end{figure*}

\begin{table*}[hbtp]
 \centering
    \begin{tabular}{l c c c}
 	\hline
    \multirow{2}{*}{Criterion}&\multirow{2}{*}{Selection}&\multirow{2}{*}{Range variation}& Variation \\ 
    & & & $\mathcal{B}(J/\psi\to\phi\eta)$ in \% \\ \hline
     $\chi_{K^+K^-\eta}^2$ &  $\chi_{K^+K^-\eta}^2 < 70$ &  $\chi_{K^+K^-\eta}^2 < 90$ & $0.7$ \\  
    $\chi^2_{\pi^+\pi^-\eta}$ & $\chi_{K^+K^-\eta}^2 < \chi^2_{\pi^+\pi^-\eta}$ & $\chi_{K^+K^-\eta}^2 < 0.8\chi^2_{\pi^+\pi^-\eta}$  & $0.5$  \\ 	 
    $M_{\gamma \gamma}$, MeV & $500 < M_{\gamma \gamma} < 600$ & $470 < M_{\gamma \gamma} < 630$ &  $0.1$ \\ 
    $M_{K^+K^-}$, MeV & $M_{K^+K^-} < 1090$ & $M_{K^+K^-} < 1110$ &  $3.4$ \\\hline
     \multicolumn{3}{c}{Sum in quadrature}  & $3.5$ \\ \hline   
    \end{tabular}
    \caption{\label{Tab:var_phi} The selection criteria variations for $\phi\eta$ process. In the fourth column the $J/\psi\to\phi\eta$ branching fraction variation from the main result is shown.}
\end{table*}

Figure \ref{fig:cuts_2} shows the (a) $\chi_{K^+K^-\eta}^2$ and the (b) $M_{\gamma\gamma}$ distributions for $\phi\eta$. The solid and dotted lines correspond to the applied selection criterion and its variation. In order to estimate the uncertainties associated with the selection criteria, Table \ref{Tab:var_phi} presents their variations. 

The branching fraction measurement correction at the 2.2\% \cite{rhopi} has been applied for the $\phi\eta$ process as in the $\rho\eta$ case.

To evaluate a systematic uncertainty associated with the non resonant events simulation, an additional simulation of the non resonant decay over the phase space is carried out. The difference between branching fraction measurements using two different simulations is taken as the non resonant decay simulation uncertainty, which results in 4.3\%. The contribution from the fixation of the number of clusters taken into account in the reconstruction is around 2.7\%. 

The fixation of the $\phi$ meson parameters such as mass and width during the simulation gives an uncertainty of around 0.6\%. The statistical contribution (associated with $\epsilon_{\phi\eta}$) due to the finite number of simulated events is around 0.5\%. The physical background is 0.2\%; it was evaluated from including processes $\rho\eta$, $K \pi K_s$, $K^+K^-\pi^0\pi^0$, $\omega\pi^0$ in the fit procedure.

Due to the significant differences between the nuclear interaction simulations for kaons using the FLUKA and GHEISHA programs, the latter is chosen for the final analysis. This choice is made because the GHEISHA simulation provided a more accurate representation of the experimental data. The uncertainty in the nuclear interaction simulation to the final result is estimated by increasing the cross section of nuclear interactions in the LKr calorimeter by 20\%. It resulted in an uncertainty of 2\%. Other sources of uncertainty are comparable to those associated with the $\rho\eta$ process (see Table~\ref{Tab:total_usr_phi_eta} for details).

 \begin{table}[hbtp]
 \centering
     \begin{tabular}{l c}
 	\hline
      Source & Uncertainty, \% \\ \hline
      Detector related uncertainty & 3.1 \\   
      Nuclear interaction & 2.0 \\ 
      Number of $J/\psi$ & 1.1\\ 	
      Simulation of non resonant $K^+K^-\eta$ & 4.3\\ 	 
      $m_{\phi}$ and $\Gamma_{\phi}$ simulation & 0.6 \\  
      $\mathcal{B}(\eta\to\gamma\gamma)$  & 0.5 \\  
      $\mathcal{B}(\phi\to K^+K^-)$  & 1.0 \\   
      Selection efficiency ($\epsilon_{\phi\eta}$)  & 0.5\\
      Physical background  & 0.2 \\ 
      Fixation of the cluster number & 2.7\\ 	 
      Selection criteria & 3.5 \\   \hline
      Total systematic uncertainty & 7.4 \\  \hline
    \end{tabular}
    \caption{\label{Tab:total_usr_phi_eta} The overall systematic uncertainties in the $\phi \eta$ process.}
\end{table}

\section{Summary}
\label{sec:conclusion}

We have presented measurements of the branching fractions for $J/\psi \to \rho\eta$ and $J/\psi \to \phi\eta$ decays using a data sample of 4.93 million $J/\psi$ events collected with the KEDR detector at the VEPP-4M collider. The results are:
\begin{align*}
\mathcal{B}(J/\psi \to \rho\eta) &= (2.04 \pm 0.58 \pm 0.39) \times 10^{-4}, \\
\mathcal{B}(J/\psi \to \phi\eta) &= (7.82 \pm 1.17 \pm 0.58) \times 10^{-4},
\end{align*}
where the first uncertainty is statistical and the second is systematic. These values show good agreement with previous measurements from MARK-III, DM2, BaBar, Belle-II and BES-II experiments.

The analysis of the $J/\psi \to \pi^+\pi^-\eta$ decay dynamics, which included continuum events, revealed possible contributions not only from the dominant $\rho\eta$ and $\omega\eta$ intermediate states, but also from $\rho(1450)\eta$ and $a_2^{\pm}\pi^{\mp}$ channels. According to this model the additional branching fractions are:
\begin{align*}
\mathcal{B}(J/\psi \to \pi^+\pi^-\eta) = (4.73 \pm 0.49 \pm 1.17) \times 10^{-4}, \\
\mathcal{B}(J/\psi \to a_2^{\pm}\pi^{\mp}) = (1.05 \pm 0.37 \pm 0.34) \times 10^{-3}, \\
\mathcal{B}(J/\psi \to \rho(1450)\eta \to \pi^+\pi^-\eta) < 1.51 \times 10^{-4}, \\\text{~at a confidence level of 90\%.}
\end{align*}

\begin{figure*}[hbtp]
\parbox{0.5\textwidth}{\centering \includegraphics [width=0.5\textwidth]{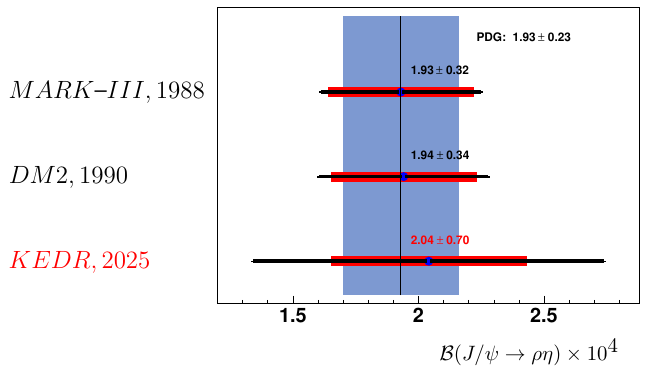}
}
\hfill
\parbox{0.5\textwidth}{\centering \includegraphics [width=0.5\textwidth]{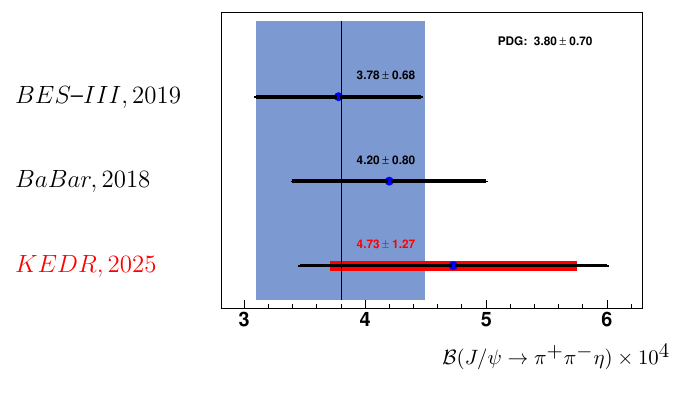}
}
\\
  \parbox{0.5\textwidth}{\centering (a)}
  \hfill
  \parbox{0.5\textwidth}{\centering (b)}
\\
 \vspace{\baselineskip}
 \\
 \parbox{1.\textwidth}{\centering \includegraphics [width=0.55\textwidth]{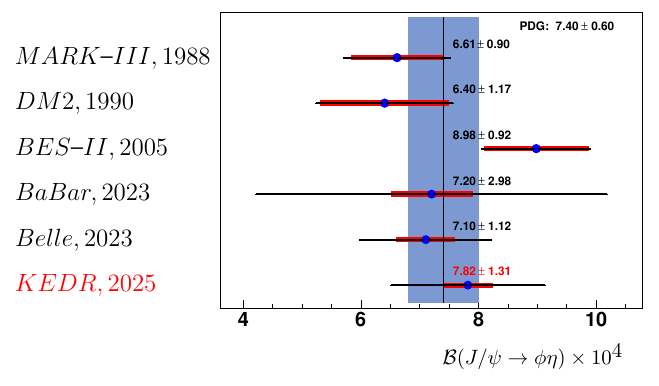}
}
 \\
  \parbox{1.\textwidth}{\centering (c)}
   \\
 \vspace{\baselineskip}
 \\
  \parbox{\textwidth}{\caption{\label{fig:branches} The comparison of the branching fraction measurements among other experiments  (a) $\rho \eta$, (b) $\pi^+ \pi^- \eta$ and (c) $\phi \eta$. Red error bars represent the systematic uncertainties, in the $\pi^+ \pi^- \eta$ case systematic uncertainties are shown only for our measurement, since the others presented the results only with a full error.}}
 \end{figure*}

Figure~\ref{fig:branches} compares the branching fraction measurements for (a) $\rho\eta$, (b) $\pi^+\pi^-\eta$ and (c) $\phi\eta$ channels with previous experiments.

Additionally, we have measured the interference phases between $\rho$-$\omega$ and $\rho$-$\rho(1450)$ resonances:

\begin{align*}
\phi_{\rho-\omega} &= (88.4 \pm 2.5 \pm 3.1)^\circ, \\
\phi_{\rho-\rho(1450)} &= (148.6 \pm 44.3 \pm 41.0)^\circ.
\end{align*}

These results provide important constraints for theoretical models of charmonium decays and will serve as valuable inputs for future studies of $J/\psi$ physics. Further improvements could be achieved with larger datasets and enhanced detector capabilities.


\section*{Acknowledgements}

We highly appreciate the VEPP-4M staff's productive work during the data collection. The authors  are grateful to A.~I.~Milstein for useful discussions of the work, as well as to the Siberian Supercomputer Center and Novosibirsk State University Supercomputer Center for providing supercomputer facilities.








\end{document}